\documentstyle[11pt,paspconf,epsf]{article}

\begin{document}

\vskip -2.0 cm
\noindent{\sl To appear in ``The Hy-Redshift Universe: \\
          Galaxy Formation and Evolution at High Redshift'' \\
          eds. A.J.~Bunker \& W.J.M.~van Breugel \\
          Astr.~Soc.~Pacif.~Conf.~Ser., in press (1999)
}
\bigskip
\bigskip

\title{
High-Redshift Quasars as Probes \\
of Galaxy and Cluster Formation
}

\author{S. G. Djorgovski}
\affil{Palomar Observatory, Caltech, Pasadena, CA 91125, USA}

\begin{abstract}
Quasars at large redshifts provide a powerful probe of structure formation
in the early universe.  Several arguments suggest that the formation of
ellipticals and massive bulges may have involved an early quasar phase.
At very large redshifts, such structures are likely to be found at the
highest peaks of the density field, and would thus be highly biased tracers:
the earliest (massive) galaxy formation may have occurred in the cores of
future rich clusters.  Preliminary results from our search for clustered
protogalaxies around quasars at $z > 4$ support this idea.  Quasars at even
larger redshifts may be an important contributor to the reionisation of the
universe, and signposts of the earliest galaxy and cluster formation.

\end{abstract}


\section{Introduction}

Hy Spinrad always hated quasars (Spinrad 1979).  In those paleolithic days
(i.e., before 1987 or so) it was not yet obvious that a powerful quasar lurks
in the heart of every one of Hy's beloved radio galaxies, but in some sense
this does not really matter: in the work of the Spinrad School of Observational
Cosmology, AGN have been used simply as means to find stellar populations at
large redshifts, in order to probe their formation and evolution.  This is in
principle a viable and sound approach.

In the simplest view, the very existence of luminous quasars at large redshifts
suggests the existence of their (massive?) host galaxies, at least in the minds 
of a vast majority of astronomers today.  At $z > 4$, this has some very
interesting and non-trivial implications for our understanding of galaxy and
structure formation (Turner 1991).

At a slightly more complex level, the observed history of the comoving number
density of quasars may be indicative of the history of galaxy formation and
evolution: the same kind of processes, i.e., dissipative mergers and tidal
interactions, may be fueling both bursts of star formation and AGN activity.
The peak seen in the comoving number density of quasars around $z \sim 2$ or 3
(Schmidt, Schneider \& Gunn 1995)
can then be interpreted in this context: the ostensible decline at high
redshifts may be indicative of the initial assembly and growth of quasar
central engines and their host galaxies; whereas the decline at lower redshifts
may be indicative of the decrease in fueling, as galaxies are carried apart
by the universal expansion, as many of the smaller pieces are being consumed,
and as the gas is being converted into stars.
Qualitatively similar predictions are made by virtually all models of
hierarchical structure formation
(see, e.g., Cataneo 1999, or Kauffmann \& Haehnelt 1999).

Due to their brightness, quasars are much easier to find (per unit telescope
time) than galaxies at comparable redshifts.
It then makes sense to use quasars as probes, or at least as pointers of sites
of galaxy formation.

Quasars have been used very effectively as probes of the intergalactic medium,
and indirectly of galaxy formation, through the studies of absorption line
systems.  A vast literature exists on this subject, which is beyond the scope
of this review; for good summaries, see, e.g., Rees (1998a) or Rauch (1998).

A good review of the searches for quasars and related topics was given by
Hartwick \& Schade (1990).  Osmer (1999) provides a modern update.  
Some of the issues covered in this review have been described by Djorgovski
(1998) and Djorgovski {\it et al.} (1999). 

\section{Quasars and Galaxy Formation}

Possibly the most direct evidence for a close relation between quasars and
galaxy formation is the remarkable correlation between the masses of central
black holes (MBH) in nearby galaxies, and the luminosities ($\sim$ masses) of
their old, metal-rich stellar populations, a.k.a. bulges (Kormendy \& Richstone
1995, Magorrian {\it et al.} 1998), with MBH's containing on average $\sim 0.6$\% of
the bulge stellar mass.  The most natural explanation for this correlation is
that both MBH's and the stellar populations are generated through a parallel
set of processes, i.e., dissipative merging and assembly at large redshifts.
Quiescent MBH's are evidently common among the normal galaxies at $z \sim 0$,
and had to originate at some point: as they grow by accretion, their formation
$is$ the quasar activity.
Quasars may thus be a common phase of the early formation of ellipticals and
massive bulges. 

Quasar demographics support this idea.  Small \& Blandford (1992), Chokshi \&
Turner (1992), and Haehnelt \& Rees (1993) all conclude that an average
$L_*$-ish galaxy today should contain an MBH with $M_\bullet \sim 10^7
~M_\odot$ or so.  These estimates (essentially integrating the known AGN
radiation over the past history of the universe) are fully consistent with the 
actual census of MBH (quasar remnants) in nearby galaxies.

Two other pieces of fossil evidence link the high-$z$ quasars with the formation
of old, metal-rich stellar populations.  First, the analysis of metallicities
in QSO BEL regions indicates super-solar abundances (up to 
$Z_Q \sim 10 ~Z_\odot$!) in quasars at $z > 4$ (Hamann \& Ferland 1993, 1999; 
Matteucci \& Padovani 1993).  The only places we know such abundances to
occur are the nuclei of giant elliptical galaxies.
Furthermore, abundance patterns in the intracluster x-ray gas at low redshifts
are suggestive of an early, rapid star formation phase in protoclusters
populated by young ellipticals (Loewenstein \& Mushotzky 1996). 

A nearly simultaneous formation of quasars and their host galaxies, or at least
ellipticals and bulges, is consistent with all of these observations, and it
fits naturally in the general picture of hierarchical galaxy and structure
formation via dissipative merging (see, e.g., 
Norman \& Scoville 1988,
Sanders {\it et al.} 1988, 
Carlberg 1990,
Hernquist \& Mihos 1995,
Mihos \& Hernquist 1996,
Monaco {\it et al.} 1999, 
Franceschini {\it et al.} 1999,
etc.).

An extreme case of this idea is that quasars are completely reducible to
ultraluminous starbursts, as advocated for many years by Terlevich and
collaborators (see, e.g., Terlevich \& Boyle 1993, and references therein).
Most other authors disagree with such a view (cf. Heckman 1991, or 
Williams \& Perry 1994), but (nearly) simultaneous manifestations of both
ultraluminous starbursts and AGN, perhaps with comparable energetics, are
clearly allowed by the data.  It is thus also possible that the early AGN
can have a profound impact on their still forming hosts, through the input
of energy and momentum (Ikeuchi \& Norman 1991, Haehnelt {\it et al.} 1998).

\section{Quasar (Proto)Clustering and Biased Galaxy Formation} 

Producing sufficient numbers of massive host galaxies needed to accommodate the
observed populations of quasars at $z > 4$, say, is not easy for most
hierarchical models: such massive halos should be rare, and associated on
average with $\sim 4$ to $5$-$\sigma$ peaks of the primordial density field 
(Efstathiou \& Rees 1988, Cole \& Kaiser 1989, Nusser \& Silk 1993).  It is a
generic prediction that for essentially every model of structure formation such
high density peaks should be strongly clustered (Kaiser 1984).  This is a purely
geometrical effect, independent of any messy astrophysical details of galaxy
formation, and thus it is a fairly robust prediction: the formation of the
first galaxies (some of which may be the hosts of high-$z$ quasars) and of the
primordial large-scale structure should be strongly coupled. 

Quasars provide a potentially useful probe of large-scale structure out to
very high redshifts.  The pre-1990 work has been reviewed by Hartwick \& Schade
(1990).  A number of quasar pairs on tens to hundreds of comoving kpc scales
has been seen (Djorgovski 1991, Kochanek {\it et al.} 1999), and some larger
groupings on scales reaching $\sim 100$ Mpc (Crampton {\it et al.} 1989, Clowes
\& Campusano 1991), but all in heterogeneous data sets.  Analysis of some more
complete samples did show a clustering signal (e.g., Iovino \& Shaver 1988,
Boyle {\it et al.} 1998).
The overall conclusion is that quasar clustering has been detected, but that
its strength decreases from $z \sim 0$ out to $z \sim 2$, the peak of the
quasar era, presumably reflecting the linear growth of the large-scale
structure.  However, if quasars are biased tracers of structure formation at
even higher redshifts, associated with very massive peaks of the primordial
density field, this trend should reverse and the clustering strength should
again start $increasing$ towards the larger look-back times. 

The first hints of such an effect were provided by the three few-Mpc quasar
pairs found in the statistically complete survey by Schneider {\it et al.}
(1994), as pointed out by Djorgovski {\it et al.} (1993) and Djorgovski (1996),
and subsequently confirmed by more detailed analysis by Kundic (1997) and 
Stephens {\it et al.} (1997).  A deeper survey for more such pairs by
Kennefick {\it et al.} (1996) did not find any more, presumably due to a
limited volume coverage.  La Franca {\it et al.} (1998) find a turn-up in
the clustering strength of quasars even at redshifts as low as $z \sim 2$.
It would be very important to check these results with new, large, complete
samples of quasars over a wide baseline in redshift.

More recently, observations of large numbers of ``field'' galaxies at 
$z \sim 3 - 3.5$ by Steidel {\it et al.} (1998) identified redshift space 
structures which are almost certainly the manifestation of biasing.
However, the effect (the bias) should be even stronger at higher redshifts,
and most of the earliest massive galaxies should be strongly clustered.
A search for protoclusters around known high-$z$ objects such as quasars thus
provides an important test of our basic ideas about the biased galaxy formation.

Intriguingly, there is a hint of a possible superclustering of quasars at
$z > 4$, on scales $\sim 100 ~h^{-1}$ comoving Mpc (cf. Djorgovski 1998). 
The effect is clearly present in the DPOSS sample (which is complete, but 
still with a patchy coverage on the sky), and in a more extended, but
heterogeneous sample of all QSOs at $z > 4$ reported to date.  The apparent
clustering in the complete sample may be an artifact of a variable depth of
the survey, which we will be able to check in a near future.  Or, it could be
due to patchy gravitational lensing magnification of the high-$z$ quasars by
the foreground large-scale structure; again, we will be able to test this
hypothesis using the DPOSS galaxy counts.  But it could also represent real
clustering of high-density peaks in the early universe, only $\sim 0.5 - 1$ Gyr
after the recombination.  The observed scale of the clustering is intriguing:
it is comparable to that corresponding to the first Doppler peak seen in CMBR
fluctuations, and to the preferred scales seen in some redshift surveys
(e.g., Broadhurst {\it et al.} 1990; Landy {\it et al.} 1996).  More data
are needed to check on this remarkable result.

\section{Quasar-Marked Protoclusters at z $>$ 4?} 

Any single search method for high-$z$ protogalaxies (PGs) has its own biases,
and formative histories of galaxies in different environments may vary 
substantially.  For example, galaxies in rich clusters are likely to start
forming earlier than in the general field, and studies of galaxy formation in
the field may have missed possible rare active spots associated with rich
protoclusters. 

We are conducting a systematic search for clustered PGs, by using quasars at $z
> 4$ as markers of the early galaxy formation sites (ostensibly protocluster
cores). The quasars themselves are selected from the DPOSS survey (Djorgovski
{\it et al.} 1999, and in prep.; Kennefick {\it et al.} 1995).  They are purely
incidental to this search: they are simply used as beacons, pointing towards
the possible sites of early, massive galaxy formation. 

\begin{figure}[t]
\plotfiddle{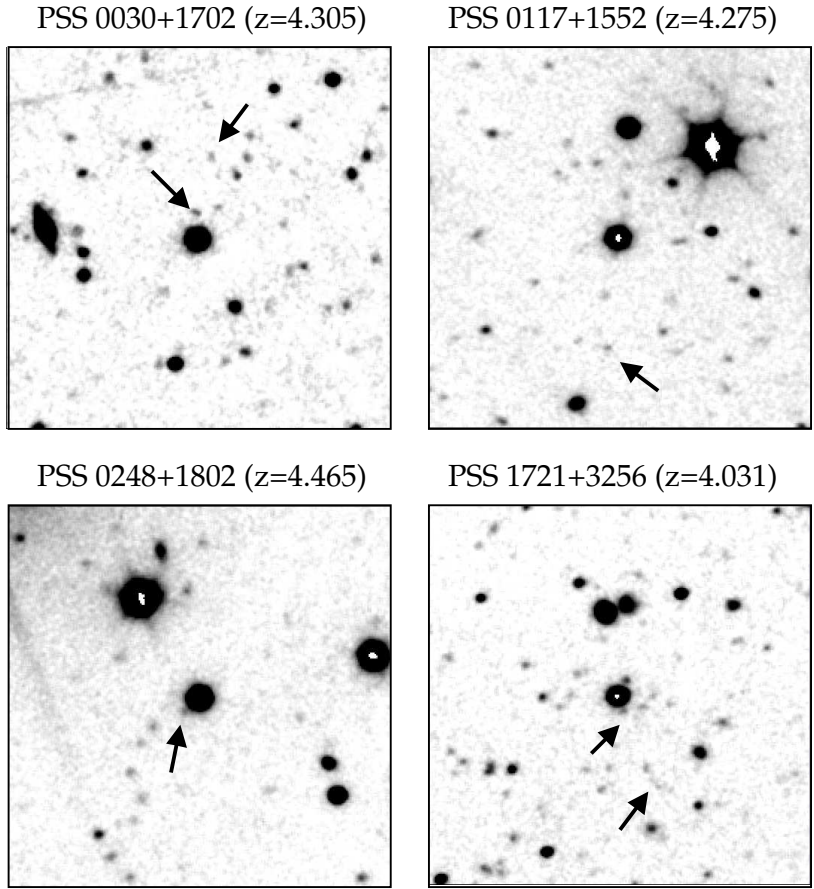}{4.5in}{0}{80}{80}{-150}{0}
\caption{ 
Examples of clustered companion protogalaxies in the fields of 4 DPOSS
quasars at $z > 4$.  These are deep, $R$-band Keck images centered on the
quasars.  The fields shown are 54 arcsec square.  Some of the spectroscopically
confirmed companions are labeled with the arrows. 
}
\end{figure}

The first galaxy discovered at $z > 3$ was a quasar companion (Djorgovski {\it
et al.} 1985, 1987). 
A Ly$\alpha$ galaxy and a dusty companion of BR 1202--0725 at $z = 4.695$ have
been discovered by several groups (Djorgovski 1995, Hu {\it et al.} 1996,
Petitjean {\it et al.} 1996), and a dusty companion object has been found in
the same field (Omont {\it et al.} 1995, Ohta {\it et al.} 1995). 
Hu \& McMahon (1996) also found two companion galaxies in the field of BR
2237--0607 at $z = 4.55$. 

We have searches to various degrees of completeness in about twenty QSO
fields so far
(Djorgovski 1998; Djorgovski {\it et al.}, in prep.).  Companion galaxies
have been found in virtually every case, despite very incomplete coverage.
They are typically located
anywhere between a few arcsec to tens of arcsec from the quasars, i.e.,
on scales $\sim 100+$ comoving kpc.
We also select candidate PGs 
by using deep $BRI$ imaging over a field of view of several arcmin,
probing $\sim 10$ comoving Mpc ($\sim$ cluster size)
projected scales.  
This is a straightforward extension of the method employed so successfully to
find the quasars themselves at $z > 4$ (at these redshifts, the continuum drop
is dominated by the Ly$\alpha$ forest, rather than the Lyman break, which is
used to select galaxies at $z \sim 2 - 3.5$).  
The candidates are followed up by multislit spectroscopy at the Keck, which
is still in progress as of this writing.

As of the mid-1999, about two dozen companion galaxies have been confirmed
spectroscopically.
Their typical magnitudes are $R \sim 25^m$, implying continuum luminosities $L
\leq L_*$.  The Ly$\alpha$ line emission is relatively weak, with typical
restframe equivalent widths $\sim 20 - 30$ \AA, an order of magnitude lower
than what is seen in quasars and powerful radio galaxies, but perfectly
reasonable for the objects powered by star formation.  There are no
high-ionization lines in their spectra,
and no signs of AGN.  The SFR inferred both from the Ly$\alpha$ line,
and the UV continuum flux is typically $\sim 5 - 10 ~M_\odot$/yr, not corrected
for the extinction, and thus it could easily be a factor of 5 to 10 times
higher. 

Overall, the intrinsic properties of these quasar companion galaxies are very
similar to those of the Lyman-break selected population at $z \sim 3 - 4$,
except of course for their special environments and somewhat higher 
look-back times. 

There is a hint of a trend that the objects closer to the quasars have stronger
Ly$\alpha$ line emission, as it may be expected due to the QSO ionization
field.  
In addition to these galaxies where we actually detect (presumably
starlight) continuum, pure Ly$\alpha$ emission line nebulae are found within
$\sim 2 - 3$ arcsec for several of the quasars, with no detectable continuum at
all.  The Ly$\alpha$ fluxes are exactly what may be expected from
photoionization by the QSO, with typical $L_{Ly \alpha} \sim$ a few $\times
10^{43}$ erg/s.  They may represent ionized parts of still gaseous protogalaxy
hosts of the quasars.  We can thus see and $distinguish$ both the objects
powered by the neighboring QSO, and ``normal'' PGs in their vicinity. 

The median projected separations of these objects from the quasars are $\sim$
a few $\times 100 h^{-1}$ comoving kpc, an order of magnitude less than the
comoving r.m.s. separation of $L_*$ galaxies today, but comparable to that in
the rich cluster cores.  The frequency of QSO companion galaxies at $z > 4$
also appears to be an order of magnitude higher than in the comparable QSO
samples at $z \sim 2 - 3$, the peak of the QSO era and the ostensible peak
merging epoch. 
However, interaction and merging rates are likely to be high in the densest
regions at high redshifts, which would naturally account for the propensity of
some of these early PGs to undergo a quasar phase, and to have close companions.

The implied average star formation density rate in these regions is some 2 or 3
orders of magnitude higher than expected from the limits estimated for these
redshifts by Madau {\it et al.} (1996) for $field$ galaxies, and 1 or 2 orders
of magnitude higher than the measurements by Steidel {\it et al.} at $z \sim
4$, even if we ignore any SFR associated with the QSO hosts (which we cannot
measure, but is surely there).  These must be very special regions of an 
enhanced galaxy formation in the early universe. 

It is also worth noting that (perhaps coincidentally) the observed comoving
number density of quasars at $z > 4$ is roughly comparable to the comoving
density of very rich clusters of galaxies today.  Of course, depending on the
timescales involved, there must be some protoclusters without observable
quasars in them, and some where more than one AGN is present (an example may be
the obscured companion of BR 1202--0725).

\section{Towards the Renaissance at z $>$ 5: the First Quasars \\
         and the First Galaxies}

The remarkable progress in cosmology over the past few years, reviewed by
several speakers at this meeting, has pushed the
frontiers of galaxy and structure formation studies out to $z > 5$.  Half a
dozen galaxies, two QSOs (cf. Fan {\it et al.} 1999), 
and one radio galaxy are now known at $z \geq 5$, 
with the most distant confirmed object at $z = 5.74$ (Hu {\it et al.} 1999).  
Remarkably, there is no convincing evidence yet for a high-$z$ decline of the
comoving star formation rate density out to $z > 4$ (Steidel {\it et al.}
1999).  Moreover, the universe at $z \sim 5$ appears to be already fully
reionised (Songaila {\it et al.} 1999, Madau {\it et al.} 1999), implying the
existence of a substantial activity in a population of sources at even higher
redshifts. 

These observational results pose something of a challenge for the models of
galaxy formation.  Essentially in all modern models, the first subgalactic
fragments with masses $\geq 10^6 ~M_\odot$ begin to form at $z \sim 10 - 30$,
and the universe becomes reionised at $z \sim 8 - 12$ (see, e.g., Gnedin \&
Ostriker 1997, Miralda-Escude \& Rees 1997, or Rauch 1998 and references
therein).  This corresponds to a time interval of only about $\sim 0.5 - 1$ Gyr
for a reasonable range of cosmologies.

What is not known is what are the first or the dominant ionisation sources
which break the ``dark ages'': primordial starbursts or primordial AGN?
This is one of the fundamental questions in cosmology today, and it dominates
many of the discussions about the NGST (see, e.g., Rees 1998b, Haiman \& Loeb
1998, or Loeb 1999).  
Optical searches for quasars at $z > 5$ have been reviewed by Osmer (1999).
There are exciting new prospects of detecting such a population in x-rays using
CXO (Haiman \& Loeb 1999).  The value of such quasars as probes of the earliest
phases of galaxy and structure formation during the reionisation era at
$z \sim 10 \times 2^{\pm 1}$ cannot be overstated.

Some numerical simulations suggest that an early formation of quasars, at
$z \sim 8$, say, is viable in the framework of the currently popular
hierarchical models with dissipation (cf. Katz {\it et al.} 1994). 
It is even possible that a substantial amount of QSO activity may predate the
peak epoch of star formation in galaxies (Silk \& Rees 1998).  
A catastrophic gravitational collapse of a massive primordial star cluster 
may be the most natural way of forming the first MBHs, but a variety of other
mechanisms have been proposed (e.g., Loeb 1993, Umemura {\it et al.} 1993,
Loeb \& Rasio 1994, etc.).  Future observations will tell whether such
primordial fireworks marked the end of the dark ages in the universe.

\acknowledgments

It is a pleasure to acknowledge the work of my collaborators, R.~Gal,
R.~Brunner, R.~de Carvalho, S.~Odewahn, and the rest of the DPOSS QSO search
team.  I also wish to thank the staff of Palomar and Keck observatories for
their expert assistance during our observing runs.  This work was supported 
in part by the Norris Foundation and by the Bressler Foundation.  Ivan King
and his LOC crew brought this meeting into existence; thank you all.
Finally, many thanks to Hy for introducing me to the joys of low-S/N astronomy
and letting me play with the big toys: it was fun (most of the time)!

\end{document}